\documentclass[11pt,ams]{article}%
\usepackage{geometry}
\usepackage{dsfont}
\usepackage{amsmath}
\usepackage{amsfonts}
\usepackage{amssymb}
\usepackage{graphicx}%

\begin{document}

\date{}
\title{\textbf{A remark on the BRST symmetry in the Gribov-Zwanziger theory }}

\author{\textbf{M.~A.~L.~Capri}$^{a}$\thanks{capri@ufrrj.br}\,\,,
\textbf{A.~J.~G\'{o}mez}$^{b}$\thanks{ajgomez@uerj.br}\,\,,
\textbf{M.~S.~Guimaraes}$^{b}$\thanks{msguimaraes@uerj.br}\,\,,
\textbf{V.~E.~R.~Lemes}$^{b}$\thanks{vitor@dft.if.uerj.br}\,\,,\\
\textbf{S.~P.~Sorella}$^{b}$\thanks{sorella@uerj.br}\ \thanks{Work supported by
FAPERJ, Funda{\c{c}}{\~{a}}o de Amparo {\`{a}} Pesquisa do Estado do Rio de
Janeiro, under the program \textit{Cientista do Nosso Estado}, E-26/101.578/2010.}\,\,,\,\,\textbf{D.~G.~Tedesco}$^{b}$\thanks{dgtedesco@uerj.br}\,\,\\[2mm]
\textit{{\small $^{a}$ UFRRJ $-$ Universidade Federal Rural do Rio de Janeiro}}\\
\textit{{\small Departamento de F\'{\i}sica $-$ Grupo de F\'{\i}sica Te\'{o}rica e Matem\'{a}tica F\'{\i}sica}}\\
\textit{{\small BR 465-07, 23890-971, Serop\'edica, RJ, Brasil.}}\\
\textit{{\small {$^{b}$ UERJ $-$ Universidade do Estado do Rio de
Janeiro}}}\\\textit{{\small {Instituto de F\'{\i}sica $-$
Departamento de F\'{\i}sica Te\'{o}rica}}}\\\textit{{\small {Rua
S{\~a}o Francisco Xavier 524, 20550-013 Maracan{\~a}, Rio de
Janeiro, RJ, Brasil.}}}$$}
\maketitle
\begin{abstract}
We show that the soft breaking of the BRST symmetry arising in the Gribov-Zwanziger theory can be converted into a linear breaking upon introduction of a set of BRST quartets of auxiliary fields. Due to its compatibility with the Quantum Action Principle, the linearly broken BRST symmetry can be directly converted into a suitable set of useful Slavnov-Taylor identities. As a consequence, it turns out that the renormalization  aspects  of the Gribov-Zwanziger theory can be addressed by means of the cohomology of a nilpotent local operator. \end{abstract}

\baselineskip=13pt

\newpage

\section{Introduction}
The Gribov-Zwanziger theory \cite{Gribov:1977wm,Zwanziger:1988jt,Zwanziger:1989mf} arises from the Landau gauge Faddeev-Popov action  when the domain of integration in the Euclidean functional integral is restricted to the so called Gribov region $\Omega$\footnote{The Gribov region $\Omega$ is defined as the set of gauge field configurations which obey the Landau gauge condition and for which the Faddeev-Popov operator is strictly positive, namely $\Omega=\{A^a_\mu, \; \partial_\mu A^a_\mu=0, \; -\partial_\mu (\partial_\mu \delta^{ab} + gf^{acb}A^c_\mu) > 0 \}$.}, whose boundary $\partial \Omega$ is known as the first Gribov horizon, such a restriction being needed in order to account for the phenomenon of the Gribov copies. So far, the Gribov-Zwanziger theory has been proven to be renormalizable \cite{Zwanziger:1988jt,Zwanziger:1989mf,Maggiore:1993wq,Dudal:2005na,Dudal:2010fq}, while providing a mechanism for gluon confinement, as displayed by the two-point gluon correlation function
\begin{equation}
\langle A^a_\mu(k) A^b_\nu(-k) \rangle = \delta^{ab} \frac{k^2}{k^4+\gamma^4} \left( \delta_{\mu\nu} -\frac{k_\mu k_\nu}{k^2} \right) \;, \label{paai}
\end{equation}
which exhibits complex poles. As such, it does not correspond to the propagation of a physical particle. The parameter $\gamma$ stands for the Gribov mass parameter. It is not a free parameter, being determined in a self-consitent way as a function of the gauge coupling constant $g$ through a gap equation, called the horizon condition \cite{Zwanziger:1988jt,Zwanziger:1989mf}.\\\\Several efforts have been undertaken in the last years \cite{Dudal:2007cw,Dudal:2008sp,Gracey:2006dr,Baulieu:2008fy,Dudal:2009xh,Sorella:2009vt,Huber:2009tx,Baulieu:2009ha,Zwanziger:2010iz}  to achieve a better understanding of the Gribov-Zwanziger theory and of its relationship with confinement. Though, there still exist  aspects of the theory which remain to be unraveled. Let  us quote, for example,  the issues of the BRST symmetry and of the construction of a set of local operators enabling us to make contact with the spectrum of Yang-Mills theories. \\\\In this  work we address the issue of the BRST symmetry. We point out that the soft breaking of the BRST symmetry exhibited by the Gribov-Zwanziger action can be converted into a linear breaking by introducing a set of BRST quartets of auxiliary fields. \\\\This observation has  far-reaching consequences. We underline that, unlike a soft breaking quadratic in the fields, a linear breaking turns out to be compatible with the Quantum Action Principle  \cite{Piguet:1995er}. The linearly broken BRST symmetry can be thus directly converted into a set of useful Slavnov-Taylor identities. Therefore, the quantum aspects of the Gribov-Zwanziger theory can be analyzed by means of the cohomology of a local nilpotent operator. In particular, both the characterization of the invariant counterterms and the renormalization of local gauge invariant composite operators can be achieved through the identification of cohomology classes of the BRST operator, for which powerful mathematical tools are available \cite{Piguet:1995er}. \\\\ Although we shall present our results within the context of the Gribov-Zwanziger action, it is worth emphasizing that the mechanism of converting the soft quadratic BRST breaking into a linear breaking equally applies to the so-called refined Gribov Zwanziger action (RGZ) introduced in \cite{Dudal:2007cw,Dudal:2008sp}. The RGZ action takes into account additional nonperturbative effects related to the existence of dimension two condensates, see \cite{Dudal:2008sp} for a detailed discussion. These condensates modify in a nontrivial way the infrared behavior of the gluon and ghost propagators. For example, the RGZ gluon propagator turns out to be \cite{Dudal:2008sp}
\begin{equation}
\langle A^a_\mu(k) A^b_\nu(-k) \rangle_{RGZ} = \delta^{ab} \frac{k^2+M^2}{k^4+(m^2+M^2)k^2 + m^2M^2+ 2g^2N\gamma^4} \left( \delta_{\mu\nu} -\frac{k_\mu k_\nu}{k^2} \right) \;, \label{rgz}
\end{equation}
where the mass parameters $M^2,m^2$ are dynamical parameters related to the nonvanishing dimension two condensates $\langle {\bar \varphi^{ab}_\mu \varphi^{ab}_\mu} -{\bar \omega^{ab}_\mu \omega^{ab}_\mu} \rangle$ and $\langle A^a_{\mu}A^a_\mu \rangle$ \cite{Dudal:2008sp}. Unlike the Gribov propagator, eq.\eqref{paai}, expression \eqref{rgz} does not vanish at zero momentum, while still providing a violation of the positivity \cite{Dudal:2008sp}, thus accounting for gluon confinement.  The infrared behavior of the gluon and ghost propagators stemming from the RGZ theory is in good agreement with the recent numerical simulations at large volume \cite{Cucchieri:2007rg,Cucchieri:2008fc,Bornyakov:2009ug}. In particular, as observed in \cite{Dudal:2010tf}, expression \eqref{rgz} provides an accurate fit of the gluon propagator up to $k\approx 1.5 GeV$. 
\\\\The present work is organized as follows. In Sect.2. a short survey on the Gribov-Zwanziger action is given. In Sect.3 we show how the soft breaking can be converted into a linear breaking upon introduction of BRST quartets. Sect.4 is devoted to the derivation of the Slavnov-Taylor identities as well as of the large set of additional Ward identities fulfilled by the new formulation of the Gribov-Zwanziger theory.

\section{A short survey on the Gribov-Zwanziger theory}

The action of the Gribov-Zwanziger theory is given by the following expression:
\begin{eqnarray}
S_{GZ} & = & \int d^4 x \left(  \frac1{4} F_{\mu\nu}^{a}F_{\mu\nu}^{a} + ib^a\partial_{\mu}A^a_\mu + {\bar c}^a \partial_\mu D_{\mu}^{ab} c^b  \right) \nonumber \\
{\ }{\ }{\ } & + &\int d^4x \left( -  {\bar \varphi}^{ac}_\mu \partial_\nu D_{\nu}^{ab} \varphi^{bc}_{\mu} +  {\bar \omega}^{ac}_\mu \partial_\nu D_{\nu}^{ab} \omega^{bc}_{\mu} + g f^{amb} (\partial_{\nu}{\bar \omega}^{ac}_{\mu}) (D^{mp}_{\nu} c^p)\varphi^{bc}_{\mu} \right) \nonumber \\
{\ }{\ }{\ } & +&   \int d^4x\left(\gamma^2\,g\,f^{abc}A_\mu^{a}(\varphi_\mu^{bc}-\bar{\varphi}_\mu^{bc})-d(N^2-1)\gamma^4 \right) \nonumber \\
& = &\frac1{4}\int d^4 x F_{\mu\nu}^{a}F_{\mu\nu}^{a} + s\int d^4 x \left(\bar{c}^a\partial_{\mu} A^a_{\mu} - \bar{\omega}_\mu^{ac}\partial_\nu D^{ab}_\nu \varphi_{\mu}^{bc}\right) + S_{\gamma} \;, \label{GZact}
\end{eqnarray}
with
\begin{align}
S_{\gamma}&=\int d^4x\left(\gamma^2\,g\,f^{abc}A_\mu^{a}(\varphi_\mu^{bc}-\bar{\varphi}_\mu^{bc})-d(N^2-1)\gamma^4 \right) \;, \label{BRSbreak}
\end{align}
where $N$ is the number of colors and $d=4$ the space-time dimension. The Gribov parameter $\gamma^2$ is determined in a self-consistent way by the horizon condition \cite{Zwanziger:1988jt,Zwanziger:1989mf}, which reads
\begin{align}
\frac{\partial {\cal E}_{vac}}{\partial \gamma^2}=0 \;,  \label{geq}
\end{align}
where $ {\cal E}_{vac} $ is the vacuum energy
\begin{align}
e^{- {\cal E}_{vac} } = \int [d\Phi] \; e^{-S_{GZ}} \;,  \label{vc}
\end{align}
and $[d\Phi]$ stands for the functional integration over all fields appearing in $S_{GZ}$. 
In the absence of the term $S_{\gamma}$, the action \eqref{GZact} enjoys the nilpotent BRST symmetry
\begin{align}
sA_\mu^{a} &= - D^{ab}_\nu c^b = -( \partial_\mu \delta^{ab} +  g f^{acb} A^c_\mu)c^b\;,\nonumber \\
sc^a &=  \frac{g}{2} f^{acb}c^b c^c\;,\nonumber\\
s\bar{c}^a &= ib^a\;,\qquad sb^a=0\;,\nonumber\\
s\bar{\omega}_\mu^{ab} &= \bar{\varphi}_\mu^{ab}\;,\qquad s\bar{\varphi}_\mu^{ab}=0\;,\nonumber\\ 
s\varphi_\mu^{ab} &= \omega_\mu^{ab}\;,\qquad s\omega_\mu^{ab}=0\;, \label{BRS}
\end{align}
and
\begin{equation}
s   \int d^4 x \left(\frac1{4} F_{\mu\nu}^{a}F_{\mu\nu}^{a} + s  \left(\bar{c}^a\partial_{\mu} A^a_{\mu} - \bar{\omega}_\mu^{ac}\partial_\nu D^{ab}_\nu \varphi_{\mu}^{bc}\right)     \right) =0  \;.
\end{equation}
The Gribov-Zwanziger action is, however, not left invariant by the BRST transformations, eqs.\eqref{BRS}, which are broken by the term $S_{\gamma}$, namely
\begin{align}
sS_{GZ} = sS_{\gamma} = \gamma^2\,\int d^4x\left(-g\,f^{abc}(D^{ad}_\mu c^d) (\varphi_\mu^{bc}-\bar{\varphi}_\mu^{bc}) +g\,f^{abc}A_\mu^{a}\omega_\mu^{bc}\right)\;. \label{BRSactbreak}
\end{align}
Notice that the breaking term, being of dimension two in the fields, is a soft breaking. Though, as the breaking is quadratic in the fields, {\it i.e.} it is a composite field operator, eq.\eqref{BRSactbreak} cannot be directly translated into a suitable set of Slavnov-Taylor identities.

\section{Converting the soft breaking into a linear breaking}
As already mentioned, it turns out that the Gribov-Zwanziger action, eq.(\ref{GZact}), admits an equivalent representation in terms of a new set of fields leading to a nilpotent BRST symmetry which is only linearly broken.  There are several ways to achieve a formulation for which the BRST symmetry is linearly broken, all relying on the introduction of a set of BRST quartets of auxiliary fields. Here, we shall present the minimal formulation in which only a pair of BRST quartets is needed. The equivalent action is given by the following expression
\begin{align}
S_{GZ}^{lin} &= \frac1{4}\int d^4 x \,F_{\mu\nu}^{a}F_{\mu\nu}^{a} + s\int d^4 x \left(\bar{c}^a\partial_{\mu} A^a_{\mu} - \bar{\omega}_\mu^{ac}\partial_\nu D^{ab}_\nu \varphi_{\mu}^{bc}\right)\nonumber\\
 &+ s\int d^4 x \left(g f^{abc} \bar{\mathcal{C}}_{\mu\nu}^{cd} A^a_{\mu}\varphi_{\nu}^{bd} - \bar{\mathcal{C}}_{\mu\nu}^{ab}\eta_{\mu\nu}^{ab}  
 -\bar{\mathcal{C}}_{\mu\nu}^{ab}\bar{\eta}_{\mu\nu}^{ab} 
 -gf^{abc} \bar{\eta}_{\mu\nu}^{cd} A^a_{\mu}\bar{\omega}_{\nu}^{bd} 
 -\bar{\rho}_{\mu\nu}^{ab}\bar{\eta}_{\mu\nu}^{ab}\right)\nonumber\\
  & +\int d^4 x \left(\gamma^2 \eta_{\mu\nu}^{ab} \delta^{ab} \delta_{\mu\nu} 
  +\gamma^2 \bar{\lambda}_{\mu\nu}^{ab} \delta^{ab} \delta_{\mu\nu}\right)\;, \label{GZactlinear}
\end{align}
where $(\bar{\mathcal{C}}_{\mu\nu}^{ab}, \lambda_{\mu\nu}^{ab}, \eta_{\mu\nu}^{ab}, \mathcal{C}_{\mu\nu}^{ab})$ and $(\bar{\rho}_{\mu\nu}^{ab}, \bar{\lambda}_{\mu\nu}^{ab}, \bar{\eta}_{\mu\nu}^{ab}, \rho_{\mu\nu}^{ab})$ are two BRST quartets of auxiliary fields, namely
\begin{align}
s\bar{\mathcal{C}}_{\mu\nu}^{ab} &=  \lambda_{\mu\nu}^{ab}\;,\qquad s\lambda_{\mu\nu}^{ab} = 0\;,\qquad 
s\eta_{\mu\nu}^{ab} =  \mathcal{C}_{\mu\nu}^{ab}\;,\qquad s\mathcal{C}_{\mu\nu}^{ab} = 0\;,\nonumber\\ 
s\bar{\rho}_{\mu\nu}^{ab} &=  \bar{\lambda}_{\mu\nu}^{ab}\;,\qquad s\bar{\lambda}_{\mu\nu}^{ab} = 0\;,\qquad
s\bar{\eta}_{\mu\nu}^{ab} =  \rho_{\mu\nu}^{ab}\;,\qquad s\rho_{\mu\nu}^{ab} = 0\;. \label{BRS2}
\end{align}
The fields $(\lambda_{\mu\nu}^{ab},\eta_{\mu\nu}^{ab})$ and $(\bar{\lambda}_{\mu\nu}^{ab}, \bar{\eta}_{\mu\nu}^{ab})$ are commuting fields, while $(\bar{\mathcal{C}}_{\mu\nu}^{ab},  \mathcal{C}_{\mu\nu}^{ab})$, $( \bar{\rho}_{\mu\nu}^{ab},  \rho_{\mu\nu}^{ab})$ are anticommuting. Each of these fields has $16(N^2-1)^2$ components\footnote{The color indices $(a,b)$ run from $1$ to $N^2-1$, while $(\mu,\nu)$ from $1$ to $4$. }.  Moreover, the fields $(\bar{\mathcal{C}}_{\mu\nu}^{ab}, \bar{\rho}_{\mu\nu}^{ab})$ have ghost number $-1$, and  $({\mathcal{C}}_{\mu\nu}^{ab}, {\rho}_{\mu\nu}^{ab})$ ghost number $1$. 
It is easily checked that, unlike expression \eqref{BRSactbreak},   the BRST symmetry defined by eqs.(\ref{BRS}) and by eqs.(\ref{BRS2}) is now linearly broken, {\it i.e.} the resulting breaking term is linear in the fields
\begin{align}
sS_{GZ}^{lin} =  \gamma^2\,\int d^4x\,\delta^{ab} \delta_{\mu\nu}\mathcal{C}_{\mu\nu}^{ab}\;. \label{BRSactbreak2}
\end{align}
In order to prove the equivalence between the two formulations,  eq.(\ref{GZact}) and eq.(\ref{GZactlinear}), we first give the explicit expression of $S_{GZ}^{lin}$ 
\begin{eqnarray}
S_{GZ}^{lin} &=& \frac{1}{4}\int d^4 x \; F_{\mu\nu}^{a}F_{\mu\nu}^{a} + s\int d^4 x \left(\bar{c}^a\partial A^a_{\mu} - \bar{\omega}_\mu^{ac}\partial_\nu D^{ab}_\nu \varphi_{\mu}^{bc}\right)\nonumber\\
 &&+\int d^4 x \left(g f^{abc} \lambda_{\mu\nu}^{cd} A^a_{\mu}\varphi_{\nu}^{bd} 
 -gf^{abc}\bar{\eta}^{cd}_{\mu\nu}\left( A^{a}_{\mu}\bar\varphi^{bd}_{\nu}-(D^{ap}_{\mu}c^{p})\bar\omega^{bd}_{\nu}\right)
 -\lambda^{ab}_{\mu\nu}\bar\eta^{ab}_{\mu\nu}\right)\nonumber\\
 &&+\int d^4 x\left(\bar{\mathcal{C}}^{cd}_{\mu\nu}\left(\mathcal{C}^{cd}_{\mu\nu}+\rho^{cd}_{\mu\nu}
 -gf^{abc}A^{a}_{\mu}\omega^{bd}_{\nu}
 +gf^{abc}(D^{ap}_{\mu}c^{p})\varphi^{bd}_{\nu}\right)
 -\rho^{cd}_{\mu\nu}(\bar\rho^{cd}_{\mu\nu}
 +gf^{abc}A^{a}_{\mu}\bar\omega^{bd}_{\nu})\right)\nonumber\\
 &&+\int d^4 x\left(\bar\lambda^{ab}_{\mu\nu}(\gamma^{2}\delta^{ab}\delta_{\mu\nu}-\bar\eta^{ab}_{\mu\nu})
 +\eta^{ab}_{\mu\nu}(\gamma^{2}\delta^{ab}\delta_{\mu\nu}-\lambda^{ab}_{\mu\nu})\right)\;. 
 \label{GZactlinear2}
\end{eqnarray}
We observe now that the new fields $(\lambda_{\mu\nu}^{ab},\eta_{\mu\nu}^{ab})$ and $(\bar{\lambda}_{\mu\nu}^{ab}, \bar{\eta}_{\mu\nu}^{ab})$ can be eliminated in an algebraic way by using their equations of motion. In addition, the anticommuting fields $(\bar{\mathcal{C}}_{\mu\nu}^{ab},  \mathcal{C}_{\mu\nu}^{ab})$, $( \bar{\rho}_{\mu\nu}^{ab},  \rho_{\mu\nu}^{ab})$ can be decoupled by suitable field redefinitions. Let us show how this works at the level of the partition function. As it is apparent from expression  \eqref{GZactlinear2}, the fields 
$\eta_{\mu\nu}^{ab}$ and $\bar{\lambda}_{\mu\nu}^{ab}$ are Lagrange multipliers. In the path integral formulation they constrain the fields $\lambda_{\mu\nu}^{ab}$ and $\bar{\eta}_{\mu\nu}^{ab}$ to take a constant value. In fact 
\begin{align}
\int [d\Xi] \; e^{-S_{GZ}^{lin}} &= \int [d\tilde{\Xi}] [d\eta][d\bar{\eta}][d\lambda][d\bar{\lambda}]\; e^{-\tilde{S}_{GZ}^{lin}  
- \int d^4 x\left(\bar\lambda^{ab}_{\mu\nu}(\gamma^{2}\delta^{ab}\delta_{\mu\nu}-\bar\eta^{ab}_{\mu\nu})
 +\eta^{ab}_{\mu\nu}(\gamma^{2}\delta^{ab}\delta_{\mu\nu}-\lambda^{ab}_{\mu\nu})\right)}\nonumber\\
  &= \int [d\tilde{\Xi}][d\bar\eta][d{\lambda}] \delta\left(\lambda_{\mu\nu}^{ab} - \gamma^2 \delta^{ab} \delta_{\mu\nu}\right)\delta\left(\bar{\eta}_{\mu\nu}^{ab} - \gamma^2 \delta^{ab} \delta_{\mu\nu}\right)\; e^{-\tilde{S}_{GZ}^{lin} }
\;,   \label{partition-delta}
\end{align}
where
\begin{eqnarray}
\tilde{S}_{GZ}^{lin} &=& \frac{1}{4}\int d^4 x\; F_{\mu\nu}^{a}F_{\mu\nu}^{a} + s\int d^4 x \left(\bar{c}^a\partial A^a_{\mu} - \bar{\omega}_\mu^{ac}\partial_\nu D^{ab}_\nu \varphi_{\mu}^{bc}\right)\nonumber\\
 &&+\int d^4 x \left(g f^{abc} \lambda_{\mu\nu}^{cd} A^a_{\mu}\varphi_{\nu}^{bd}
 -gf^{abc}\bar{\eta}^{cd}_{\mu\nu} \left( A^{a}_{\mu}\bar\varphi^{bd}_{\nu}-(D^{ap}_{\mu}c^{p})\bar\omega^{bd}_{\nu}\right)
 -\lambda^{ab}_{\mu\nu}\bar\eta^{ab}_{\mu\nu}\right)\nonumber\\
 &&+\int d^4 x\left(\bar{\mathcal{C}}^{cd}_{\mu\nu}\left( \mathcal{C}^{cd}_{\mu\nu}+\rho^{cd}_{\mu\nu}
 -gf^{abc}A^{a}_{\mu}\omega^{bd}_{\nu}
 +gf^{abc}(D^{ap}_{\mu}c^{p})\varphi^{bd}_{\nu} \right)
 -\rho^{cd}_{\mu\nu}(\bar\rho^{cd}_{\mu\nu}
 +gf^{abc}A^{a}_{\mu}\bar\omega^{bd}_{\nu})\right)\; \label{GZactlinear3}
\end{eqnarray}
Performing the integration over $\lambda_{\mu\nu}^{ab}$ and $\bar{\eta}_{\mu\nu}^{ab}$, it follows 
\begin{equation}
\int [d\Xi] \; e^{-S_{GZ}^{lin}}  = \int [d\tilde{\Xi}]\; e^{-{\hat{S}}_{GZ}^{lin}} \;, \label{eqa1}
\end{equation} 
with 
\begin{equation}
{\hat{S}}_{GZ}^{lin} = \tilde{S}_{GZ}^{lin} \big|_{  \lambda_{\mu\nu}^{ab} = \gamma^2 \delta^{ab} \delta_{\mu\nu} ;  \bar{\eta}_{\mu\nu}^{ab} = \gamma^2 \delta^{ab} \delta_{\mu\nu}}  \;, \label{eqa2}
\end{equation}
where $\Xi$ is a shorthand notation to denote all fields appearing in $S_{GZ}^{lin}$, while $\tilde{\Xi}$ refers to all fields of ${\hat{S}}_{GZ}^{lin}$, {\it i.e.} it does not contain $\eta,\bar{\eta},\lambda,\bar{\lambda}$. \\\\The action ${\hat{S}}_{GZ}^{lin}$, eq.\eqref{eqa2}, takes the form
\begin{eqnarray}
{\hat{S}}_{GZ}^{lin} &=& S_{GZ}
+\int d^4 x\;\gamma^{2}gf^{abc}(D^{ad}_{\mu}c^{d})\bar\omega^{bc}_{\mu}\nonumber\\
 &+&\int d^4 x\;\bar{\mathcal{C}}^{cd}_{\mu\nu}\left( \mathcal{C}^{cd}_{\mu\nu}+\rho^{cd}_{\mu\nu}
 -gf^{abc}A^{a}_{\mu}\omega^{bd}_{\nu}
 +gf^{abc}(D^{ap}_{\mu}c^{p})\varphi^{bd}_{\nu} \right) \nonumber\\
 &-&\int d^4 x\;\rho^{cd}_{\mu\nu}(\bar\rho^{cd}_{\mu\nu}
 +gf^{abc}A^{a}_{\mu}\bar\omega^{bd}_{\nu})\;.  \label{GZequiv}
\end{eqnarray}
We proceed now by first redefining the variables  ${\mathcal{C}}_{\mu\nu}^{cd}, \bar\rho^{cd}_{\nu}$ as 
\begin{align}
\tilde{\mathcal{C}}_{\mu\nu}^{cd} &\equiv \mathcal{C}_{\mu\nu}^{cd}
 +\rho^{cd}_{\mu\nu}
 -gf^{abc}A^{a}_{\mu}\omega^{bd}_{\nu}
 +gf^{abc}(D^{ap}_{\mu}c^{p})\varphi^{bd}_{\nu}\;, \nonumber\\
\tilde{\bar\rho}_{\mu\nu}^{cd} &\equiv \bar\rho^{cd}_{\nu}
 +gf^{abc}A^{a}_{\mu}\bar\omega^{bd}_{\nu}\;,  \label{GZequiv2a}
\end{align}
which has the effect of decoupling them from the action. Further, we eliminate the term $\gamma^{2}gf^{abc}(D^{ad}_{\mu}c^{d})\bar\omega^{bc}_{\mu}$ in expression \eqref{GZequiv} by redefining the field $\omega^{ab}_{\mu}$ as\footnote{It is useful to pint out that the redefinition \eqref{GZequiv2b}, albeit nonlocal due to the presence of $ [(\partial\cdot{D})^{-1}]$, is perfectly allowed within the Gribov region $\Omega$, in which the operator $-\partial_\mu D_\mu$ is strictly positive.}
\begin{align}
\tilde{\omega}^{bc}_{\mu}&\equiv \omega^{bc}_{\mu}+[(\partial\cdot{D})^{-1}]^{bd}\left( \gamma^{2}gf^{dec}D^{ep}_{\mu}c^{p} \right)\;,\nonumber\\
  \label{GZequiv2b}
\end{align}
As the redefinitions \eqref{GZequiv2a}, \eqref{GZequiv2b} have unity Jacobian, it follows
\begin{align}
\int [d\tilde{\Xi}]\;  e^{{\hat{S}}_{GZ}^{lin}  }  &= \int [d\Phi][d\bar{\mathcal{C}}][d\tilde{\mathcal{C}}][d\tilde{\bar{\rho}}][d{\rho}] e^{-S_{GZ} -  \int d^4 x\; \left(\bar{\mathcal{C}}_{\mu\nu}^{ab} \tilde{\mathcal{C}}_{\mu\nu}^{ab}  + \tilde{\bar{\rho}}_{\mu\nu}^{ab}{\rho}_{\mu\nu}^{ab} \right)  } \nonumber\\
 &=  {\cal N}\int [d\Phi] e^{-S_{GZ} }   \;,  \label{GZequiv3}
\end{align}
where ${\cal N}$ is a constant factor.  Expression (\ref{GZequiv3})  shows thus  the equivalence between $S_{GZ}$, given by eq.(\ref{GZact}) and $S_{GZ}^{lin}$, given by eq.(\ref{GZactlinear}).

\section{The Ward identities}
The linearly broken BRST symmetry, eq.\eqref{BRSactbreak2}, can be directly converted into a useful set of Slavnov-Taylor identities. This stems from the fact that an equation of the type of \eqref{BRSactbreak2} turns out to be compatible with the Quantum Action Principle \cite{Piguet:1995er}. In order to derive the Slavnov-Taylor identities, it is useful to follow \cite{Zwanziger:1988jt,Zwanziger:1989mf,Maggiore:1993wq,Dudal:2005na,Dudal:2010fq}  and introduce a multi-index notation
\begin{eqnarray}
\left( \varphi_{\mu}^{ab}, \bar{\varphi}_{\mu}^{ab}, \omega_{\mu}^{ab}, \bar{\omega}_{\mu}^{ab} \right) &= \left( \varphi_{i}^{a}, \bar{\varphi}_{i}^{a}, \omega_{i}^{a}, \bar{\omega}_{i}^{a} \right)\;, \nonumber\\
\left( \bar{\mathcal{C}}_{\mu\nu}^{ab}, \lambda_{\mu\nu}^{ab}, \eta_{\mu\nu}^{ab}, \mathcal{C}_{\mu\nu}^{ab} \right) &= \left(\bar{\mathcal{C}}_{\mu i}^{a}, \lambda_{\mu i}^{a}, \eta_{\mu i}^{a}, \mathcal{C}_{\mu i}^{a} \right)\;, \nonumber\\
\left( \bar{\rho}_{\mu\nu}^{ab}, \bar{\lambda}_{\mu\nu}^{ab}, \bar{\eta}_{\mu\nu}^{ab}, \rho_{\mu\nu}^{ab} \right) &= \left(\bar{\rho}_{\mu i}^{a}, \bar{\lambda}_{\mu i}^{a}, \bar{\eta}_{\mu i}^{a}, \rho_{\mu i}^{a} \right)   \;,  \label{multiindex}
\end{eqnarray}
where $i = 1,...,f$, with $f=d(N^2-1)$. Therefore 
\begin{eqnarray}
S_{GZ}^{lin} &=& \int d^4 x \left(  \frac1{4} F_{\mu\nu}^{a}F_{\mu\nu}^{a} + ib^a\partial_{\mu}A^a_\mu + {\bar c}^a \partial_\mu D_{\mu}^{ab} c^b  \right) \nonumber \\
{\ }{\ }{\ } & + &\int d^4x \left( -  {\bar \varphi}^{a}_i \partial_\nu D_{\nu}^{ab} \varphi^{b}_i +  {\bar \omega}^{a}_i \partial_\nu D_{\nu}^{ab} \omega^{b}_i + g f^{amb} (\partial_{\nu}{\bar \omega}^{a}_i) (D^{mp}_{\nu}c^p) \varphi^{b}_i \right) \nonumber \\
&+& \int d^4 x \left( g f^{abc} \lambda_{\mu i}^{c} A^a_{\mu}\varphi_i^{b} 
 -gf^{abc}\bar{\eta}^{c}_{\mu i}\left( A^{a}_{\mu}\bar\varphi^{b}_{i}-(D^{ap}_{\mu}c^{p})\bar\omega^{b}_{i} \right)
 -\lambda^{a}_{\mu i}\bar\eta^{a}_{\mu i} \right) \nonumber\\
 &+&\int d^4 x \left( {\bar{\mathcal{C}}^{c}_{\mu i}}  \left( {\mathcal{C}}^{c}_{\mu i}+\rho^{c}_{\mu i}
 -gf^{abc}A^{a}_{\mu}\omega^{b}_{i}
+gf^{abc}(D^{ap}_{\mu}c^{p}) \varphi^{b}_{i} \right) 
 -\rho^{c}_{\mu i}(\bar\rho^{c}_{\mu i}
 +gf^{abc}A^{a}_{\mu}\bar\omega^{b}_{i})  \right)  \nonumber\\
 &+&\int d^4 x\left(\bar\lambda^{a}_{\mu i}(\gamma^{2}\delta^{ab}\delta_{\mu\nu} \delta^{i}_{\nu b}-\bar\eta^{a}_{\mu i})
 +\eta^{a}_{\mu i}(\gamma^{2}\delta^{ab}\delta_{\mu\nu}\delta^{i}_{\nu b}-\lambda^{a}_{\mu i})\right)\;. 
 \label{GZactlinear2_i}
\end{eqnarray}
Introducing thus two BRST invariant external sources $(\Omega^a_\mu, L^a)$ coupled to the nonlinear BRST variations of the fields $(A^a_\mu,c^a)$ 
\begin{align}
\Sigma_{GZ}^{lin} = S_{GZ}^{lin} +  \int d^4 x\; \left( - \Omega_{\mu}^{a} D^{ab}_\mu c^b + \frac{g}{2} f^{acb}L^a c^b c^c\right) \;,  \label{quant-GZlinear}
\end{align}
it follows that the action $\Sigma_{GZ}^{lin}$ fulfills the linearly broken Slavnov-Taylor identities
\begin{align}
\mathcal{S}(\Sigma_{GZ}^{lin}) =  \gamma^2\,\int d^4x\;\delta^{ab} \delta_{\mu\nu}\delta_{b\nu}^i \mathcal{C}_{\mu i}^{a} \label{st1}
\end{align}
where 
\begin{align}
\mathcal{S}(\Sigma_{GZ}^{lin})&= \int d^{4}x\,\biggl( \frac{\delta\Sigma_{GZ}^{lin}}{\delta A_{\mu}^a}\frac{\delta\Sigma_{GZ}^{lin}}{\delta \Omega_{\mu}^a} + \frac{\delta\Sigma_{GZ}^{lin}}{\delta L^a}\frac{\delta\Sigma_{GZ}^{lin}}{\delta c^a}+ ib^a\frac{\delta\Sigma_{GZ}^{lin}}{\delta \bar{c}^a} 
  + \lambda_{\mu i}^a\frac{\delta\Sigma_{GZ}^{lin}}{\delta \bar{\mathcal{C}}_{\mu i}^a}+ \mathcal{C}_{\mu i}^a\frac{\delta\Sigma_{GZ}^{lin}}{\delta \eta_{\mu i}^a} \nonumber\\
  &+ \bar{\lambda}_{\mu i}^a\frac{\delta\Sigma_{GZ}^{lin}}{\delta \bar{\rho}_{\mu i}^a} + \rho_{\mu i}^a\frac{\delta\Sigma_{GZ}^{lin}}{\delta \bar{\eta}_{\mu i}^a} + \bar{\varphi}^{a}_{i}\frac{\delta\Sigma}{\delta\bar{\omega}^{a}_{i}}
+ \omega^{a}_{i}\frac{\delta\Sigma}{\delta\varphi^{a}_{i}}\biggr) \, \label{sti}
\end{align} 
From eq.\eqref{st1} it follows that the linearized Slavnov-Taylor operator, defined as 
\begin{eqnarray} 
{\cal B}_{\Sigma_{GZ}^{lin}} & = & \int d^{4}x\, \biggl( \frac{\delta\Sigma_{GZ}^{lin}}{\delta A_{\mu}^a}\frac{\delta }{\delta \Omega_{\mu}^a} + \frac{\delta\Sigma_{GZ}^{lin}}{\delta \Omega_{\mu}^a}\frac{\delta }{\delta A_{\mu}^a}
+ \frac{\delta\Sigma_{GZ}^{lin}}{\delta L^a}\frac{\delta }{\delta c^a} +
\frac{\delta\Sigma_{GZ}^{lin}}{\delta c^a}\frac{\delta }{\delta L^a}  
+  ib^a\frac{\delta }{\delta \bar{c}^a}  \nonumber \\
&{\ }{\ }{\ }{\ }{\ }{\ }& +   \lambda_{\mu i}^a\frac{\delta}{\delta \bar{\mathcal{C}}_{\mu i}^a}+ \mathcal{C}_{\mu i}^a\frac{\delta}{\delta \eta_{\mu i}^a}+ \bar{\lambda}_{\mu i}^a\frac{\delta}{\delta \bar{\rho}_{\mu i}^a} + \rho_{\mu i}^a\frac{\delta}{\delta \bar{\eta}_{\mu i}^a}+ \bar{\varphi}^{a}_{i}\frac{\delta}{\delta\bar{\omega}^{a}_{i}}
+ \omega^{a}_{i}\frac{\delta}{\delta\varphi^{a}_{i}} \biggr) \, \label{linsti}
\end{eqnarray} 
is nilpotent, {\it i.e.} 
\begin{equation} 
{\cal B}_{\Sigma_{GZ}^{lin}} {\cal B}_{\Sigma_{GZ}^{lin}}= 0 \;. \label{nilp}
\end{equation}
We remind here that, according to the framework of the algebraic renormalization \cite{Piguet:1995er}, the invariant counterterms needed to renormalize the theory correspond to the cohomology of the linearized operator ${\cal B}_{\Sigma_{GZ}^{lin}}$ in the space of the integrated local polynomials in the fields with dimensions bounded by four. \\\\In addition to the Slavnov-Taylor identities, eq.\eqref{sti}, the action ${\Sigma_{GZ}^{lin}}$ fulfills a rather large set of additional Ward identities, which we enlist below: 
\begin{enumerate}
  \item The equations of motion of the fields $b^a$, $\bar{c}^a$, $\mathcal{C}_{\mu i}^{a}$, $\bar\rho_{\mu i}^{a}$, $\eta_{\mu i}^{a}$ and $\bar{\lambda}_{\mu i}^{a}$
\begin{equation}
\frac{\delta \Sigma_{GZ}^{lin}}{\delta b^a} = i \partial_{\mu} A_{\mu}^a  \;, \qquad
\frac{\delta \Sigma_{GZ}^{lin}}{\delta \bar{c}^a} + \partial_{\mu}\frac{\delta \Sigma_{GZ}^{lin}}{\delta \Omega_{\mu}^a}  = 0 \;, \label{eqmotion1}
\end{equation}
\begin{equation} 
\frac{\delta \Sigma_{GZ}^{lin}}{\delta \eta_{\mu i}^{a}} =  -\lambda_{\mu i}^{a} + \gamma^2 \delta^{ab} \delta_{\mu\nu}\delta_{b\nu}^i  \;, \qquad 
\frac{\delta \Sigma_{GZ}^{lin}}{\delta \bar{\lambda}_{\mu i}^{a}} = -\bar{\eta}_{\mu i}^{a} + \gamma^2 \delta^{ab} \delta_{\mu\nu}\delta_{b\nu}^i    \;, \label{eqmotion2}
\end{equation}
\begin{equation} 
\frac{\delta \Sigma_{GZ}^{lin}}{\delta \mathcal{C}_{\mu i}^{a}} = -\bar{\mathcal{C}}_{\mu i}^{a}  \;, \qquad
\frac{\delta \Sigma_{GZ}^{lin}}{\delta \bar\rho_{\mu i}^{a}} = {\rho}_{\mu i}^{a}  \;. \label{eqmotion3}
\end{equation}
Notice that all breakings appearing in eqs.\eqref{eqmotion1}, \eqref{eqmotion2}, \eqref{eqmotion3} are linear in the fields.

 \item The Ward identity for the Gribov parameter $\gamma$, namely 
\begin{align}
\frac{\partial \Sigma_{GZ}^{lin}}{\partial \gamma^2} = \int d^4x\;\delta^{ab} \delta_{\mu\nu}\delta_{b\nu}^i (\eta_{\mu i}^{a}+\bar{\lambda}_{\mu i}^{a}) \;. \label{nonrenorm}
\end{align}
Again, this identity exhibits a linear breaking. We underline that the Ward identity \eqref{nonrenorm} has a very special role, as it enables us to control the dependence of the invariant counterterms from the Gribov parameter $\gamma$. In particular, this equation provides a simple understanding of the nonrenormalization properties enjoyed by the Gribov parameter, as already reported in \cite{Zwanziger:1988jt,Zwanziger:1989mf,Maggiore:1993wq,Dudal:2005na,Dudal:2010fq}.   

  \item The local, linearly broken, equation of motion of $\bar{\varphi}_i^a$
\begin{align}
\frac{\delta \Sigma_{GZ}^{lin}}{\delta \bar{\varphi}_i^a} 
+\partial_{\mu}\frac{\delta \Sigma_{GZ}^{lin}}{\delta \lambda_{\mu i}^a}
+gf^{abc}A_{\mu}^b\frac{\delta \Sigma_{GZ}^{lin}}{\delta \bar{\lambda}_{\mu i}^c} 
= \Delta^{ai}_{\bar{\varphi}} \;, \label{eqmotion-barphi}
\end{align}
where
\begin{align}
\Delta^{ai}_{\bar{\varphi}} = -\partial^2 \varphi_i^a - \partial_{\nu}\eta_{\nu i}^a - \partial_{\mu}\bar{\eta}_{\mu i}^a -  g\gamma^2 f^{abc}A_{\mu}^c\delta_{b\mu}^i\;. \label{delta-barphi}
\end{align}

 \item The local, linearly broken, equation of motion of $\bar{\omega}_i^a$
\begin{align}
\frac{\delta \Sigma_{GZ}^{lin}}{\delta \bar{\omega}_i^a} 
+\partial_{\mu}\frac{\delta \Sigma_{GZ}^{lin}}{\delta \bar{\mathcal{C}}_{\mu i}^a} 
+gf^{abc}A^{b}_{\mu}\frac{\delta\Sigma_{GZ}^{lin}}{\delta\bar\rho^{c}_{\mu i}}
+ gf^{abc}\left(\frac{\delta\Sigma_{GZ}^{lin}}{\delta\bar\lambda^{b}_{\mu i}}-\gamma^{2}\delta^{i}_{\mu b}\right)
\frac{\delta\Sigma_{GZ}^{lin}}{\delta\Omega^{c}_{\mu}}
= \Delta^{ai}_{\bar{\omega}}  \;, \label{eqmotion-baromega}
\end{align}
where
\begin{align}
\Delta^{ai}_{\bar{\omega}} = \partial^2 \omega_i^a + \partial_{\mu}\mathcal{C}_{\mu i}^a
+\partial_{\mu}\rho^{a}_{\mu i} \;. \label{delta-baromega}
\end{align}

\item The local, linearly broken, equation of motion of $\varphi_i^a$
\begin{align}
\frac{\delta \Sigma_{GZ}^{lin}}{\delta \varphi_i^a} 
-\partial_{\mu}\frac{\delta\Sigma_{GZ}^{lin}}{\delta\bar\eta^{a}_{\mu i}}
-igf^{abc}\bar\varphi^{b}_{i}\frac{\delta\Sigma_{GZ}^{lin}}{\delta b^{c}}
+gf^{abc}\bar\omega^{b}_{i}\frac{\delta\Sigma_{GZ}^{lin}}{\delta\bar{c}^{c}}
-gf^{abc}A^{b}_{\mu} \frac{\delta\Sigma_{GZ}^{lin}}{\delta\eta^{c}_{\mu i}} 
-gf^{acm} \frac{\delta\Sigma_{GZ}^{lin}}{\delta\mathcal{C}^{c}_{\mu i}}
\frac{\delta\Sigma_{GZ}^{lin}}{\delta\Omega_{\mu}^m}
= \Delta^{ai}_\varphi \;, \label{eqmotion-phi}
\end{align}
where
\begin{align}
\Delta^{ai}_\varphi =-\partial^{2}\bar\varphi^{a}_{i}
+\partial_{\mu}\lambda^{a}_{\mu i}
+\partial_{\mu}\bar\lambda^{a}_{\mu i} 
-\gamma^2gf^{abc}A_{\mu}^b\delta_{c\mu}^i\;. \label{delta-phi}
\end{align}

\item The local, linearly broken,  equation of motion of $\omega_i^a$
\begin{align}
  \frac{\delta \Sigma_{GZ}^{lin}}{\delta \omega_i^a} 
  -\partial_{\mu}\frac{\delta\Sigma_{GZ}^{lin}}{\delta\rho^{a}_{\mu i}}
   -igf^{abc}\bar{\omega}_i^b\frac{\delta \Sigma_{GZ}^{lin}}{\delta b^c} 
   -gf^{abc}A_{\mu}^b\frac{\delta \Sigma_{GZ}^{lin}}{\delta {\mathcal{C}}_{\mu i}^c} = \Delta^{ai}_{\omega}\;, \label{eqmotion-omega}
\end{align}
where
\begin{align}
\Delta^{ai}_{\omega}=-\partial^{2}\bar\omega^{a}_{i}+\partial_{\mu}\bar\rho^{a}_{\mu i} + \partial_{\mu} \bar{\mathcal{C}}^a_{\mu i}  \;.
\end{align}

\item The integrated Ward identity
\begin{align}
\int d^4x\; \left(c^a\frac{\delta \Sigma_{GZ}^{lin}}{\delta \omega_i^a} - \bar{\omega}_i^a\frac{\delta \Sigma_{GZ}^{lin}}{\delta \bar{c}^a} + \frac{\delta\Sigma_{GZ}^{lin}}{\delta \mathcal{C}_{\mu i}^c}\frac{\delta\Sigma_{GZ}^{lin}}{\delta \Omega_{\mu}^c} -\partial_{\mu}c^c \frac{\delta\Sigma_{GZ}^{lin}}{\delta \mathcal{C}_{\mu i}^c}\right)= 0\;. \label{ward-ident}
\end{align}

\item The ghost equation \cite{Piguet:1995er,Blasi:1990xz}
\begin{align}
\mathcal{G}^{a}(\Sigma_{GZ}^{lin})=\Delta^{a}_{c}\;,
\end{align}
where
\begin{align}
\mathcal{G}^{a}&=\int d^{4}x\biggl[\frac{\delta}{\delta{c}^{a}}
+gf^{abc}\biggl(-i\bar{c}^{b}\frac{\delta}{\delta b^{c}}
+\bar\omega^{b}_{i}\frac{\delta}{\delta\bar\varphi^{c}_{i}}
+\varphi^{b}_{i}\frac{\delta}{\delta\omega^{c}_{i}}
+\bar\eta^{b}_{\mu i}\frac{\delta}{\delta\rho^{c}_{i}}
+\bar{\mathcal{C}}^{b}_{\mu i}\frac{\delta}{\delta\lambda^{c}_{\mu i}}
+\bar\rho^{b}_{\mu i}\frac{\delta}{\delta\bar\lambda^{c}_{\mu i}}
+\eta^{b}_{\mu i}\frac{\delta}{\delta\mathcal{C}^{c}_{\mu i}}\biggr)\biggr]\;,
\end{align}
and $\Delta^{a}_{c}$  is a linear breaking
\begin{align}
\Delta^{a}_{c}=\int d^{4}x\,gf^{abc}(\Omega^{b}_{\mu}A^{c}_{\mu}
-L^{b}c^{c}
-\gamma^{2}\delta^{i}_{c\mu}\bar\rho^{b}_{\mu i})\;.
\end{align}
\item The linearly broken identity
\begin{equation}
{\cal N}_{ij} \left(\Sigma_{GZ}^{lin}\right)  = -\gamma^2 \int d^4x\; \delta^{ab} \delta_{\mu\nu} \delta^{j}_{\nu b} \left( {\bar \rho}^a_{\mu i}  +  \bar{\mathcal{C}}^{a}_{\mu i}    \right)  \;, \label{nij}
\end{equation}
and 
\begin{equation}
{\cal N}_{ij}  = \int d^4x \left(   - {\bar \omega}^a_i \frac{\delta}{\delta {\bar \varphi}^a_j} +  {\varphi}^a_j \frac{\delta}{\delta {\omega}^a_i} +{\bar \eta}^a_{\mu j} \frac{\delta}{\delta {\rho}^a_{\mu i}}  
-{\bar{\mathcal{C}}}^a_{\mu i} \frac{\delta}{\delta {\lambda}^a_{\mu j}}    
- ({\bar{\rho}^a_{\mu i}}+ {\bar{\mathcal{C}}}^a_{\mu i} )\frac{\delta}{\delta {\bar{\lambda}}^a_{\mu j}}     
+( {\eta}^a_{\mu j} + {\bar \eta}^a_{\mu j}) \frac{\delta}{\delta {\cal C}^a_{\mu i}}  
\right)  \label{nijop}
\end{equation}
\item The linearly broken global symmetry $U(f)$
\begin{align}
\mathcal{Q}_{ij}(\Sigma_{GZ}^{lin})=    \gamma^2 \delta^{ab} \delta_{\mu\nu}\int d^4x\;(\delta_{b\nu}^j \eta_{\mu i}^{a}-\delta_{b\nu}^i \bar{\lambda}_{\mu j}^{a})    \;,
\end{align}
where
\begin{align}
\mathcal{Q}_{ij}&=\int d^4x\ \left(\varphi_i^a\frac{\delta }{\delta \varphi_j^a} 
-\bar{\varphi}_j^a\frac{\delta }{\delta \bar{\varphi}_i^a} 
+\omega_i^a\frac{\delta }{\delta \omega_j^a} 
-\bar{\omega}_i^a\frac{\delta}{\delta \bar{\omega}_j^a}  
+\mathcal{C}_{\mu i}^a\frac{\delta }{\delta \mathcal{C}_{\mu j}^a}
-\bar{\mathcal{C}}_{\mu j}^a\frac{\delta }{\delta \bar{\mathcal{C}}_{\mu i}^a}
\right.\nonumber\\ 
&\left.
+\rho_{\mu i}^a\frac{\delta}{\delta \rho_{\mu j}^a}
-\bar{\rho}_{\mu j}^a\frac{\delta }{\delta \bar{\rho}_{\mu i}^a}
+\eta_{\mu i}^a\frac{\delta}{\delta \eta_{\mu j}^a} 
-\lambda_{\mu j}^a\frac{\delta }{\delta \lambda_{\mu i}^a}
+\bar{\eta}_{\mu i}^a\frac{\delta}{\delta \bar{\eta}_{\mu j}^a}
-\bar{\lambda}_{\mu j}^a\frac{\delta }{\delta \bar{\lambda}_{\mu i}^a}
 \right)\;, 
\label{uf}
\end{align}
\item The linearly broken rigid identity 
\begin{equation} 
{\cal W}^a (\Sigma_{GZ}^{lin}) = \gamma^2 f^{abc} \int d^4x\; \left(    {\bar \lambda}^{bc}_{\mu\mu} +  {\eta}^{bc}_{\mu\mu} \right) \;, \label{color}
\end{equation}
with
\begin{eqnarray} 
{\cal W}^a  &=& \int d^4x\; f^{abc}\Bigl( A^b_\mu \frac{\delta}{\delta A^c_\mu} +\Omega^b_\mu \frac{\delta}{\delta \Omega^c_\mu} +c^b \frac{\delta}{\delta c^c} +L^b \frac{\delta}{\delta L^c} +{\bar c}^b \frac{\delta}{\delta {\bar c}^c} +b^b \frac{\delta}{\delta b^c}  + {\bar \omega}^b_i \frac{\delta}{\delta {\bar \omega}^c_i} + { \omega}^b_i \frac{\delta}{\delta {\omega}^c_i}  + {\bar \varphi}^b_i \frac{\delta}{\delta {\bar \varphi}^c_i}  \nonumber \\
&+& {\varphi}^b_i \frac{\delta}{\delta {\varphi}^c_i} +{\bar \eta}^b_{\mu i} \frac{\delta}{\delta {\bar \eta}^c_{\mu i} }
+{ \eta}^b_{\mu i} \frac{\delta}{\delta { \eta}^c_{\mu i} } +{\bar{\mathcal{C}}}^b_{\mu i} \frac{\delta}{\delta {\bar{\mathcal{C}}}^c_{\mu i} } + {{\mathcal{C}}}^b_{\mu i} \frac{\delta}{\delta {{\mathcal{C}}}^c_{\mu i} }
+{\bar{\rho}}^b_{\mu i} \frac{\delta}{\delta {\bar{
\rho}}^c_{\mu i} }
+{\rho}^b_{\mu i} \frac{\delta}{\delta {\rho}^c_{\mu i} }
+{\bar{\lambda}}^b_{\mu i} \frac{\delta}{\delta {\bar{
\lambda}}^c_{\mu i} }
+{\lambda}^b_{\mu i} \frac{\delta}{\delta {\lambda}^c_{\mu i} }
\Bigl) \nonumber \\
\label{colorop}
\end{eqnarray}

\end{enumerate}

\section{Conclusion} 

In this  work the issue of the BRST symmetry in the Gribov-Zwanziger theory has been addressed. We have pointed out that the soft breaking of the BRST symmetry exhibited by the Gribov-Zwanziger action can be converted into a linear breaking upon introduction of a set of BRST quartets of auxiliary fields. Due to its compatibility with the Quantum Action Principle \cite{Piguet:1995er}, the linearly broken BRST symmetry gives rise to suitable Slavnov-Taylor identities, as summarized by eq.\eqref{st1}. The renormalization aspects of the theory can thus be addressed by looking at the cohomology of the nilpotent linearized operator ${\cal B}_{\Sigma_{GZ}^{lin}}$, eq.\eqref{linsti}. \\\\Although the details of the renormalizability of the Gribov-Zwanziger theory in the new set of variables will be reported in a more detailed work, 
we believe that the present observation might improve our current understanding of the issue of the BRST symmetry in the presence of the Gribov horizon. 

\section*{Acknowledgments}
The Conselho Nacional de Desenvolvimento Cient\'{\i}fico e
Tecnol\'{o}gico (CNPq-Brazil), the Faperj, Funda{\c{c}}{\~{a}}o de
Amparo {\`{a}} Pesquisa do Estado do Rio de Janeiro, the Latin
American Center for Physics (CLAF), the SR2-UERJ,  the
Coordena{\c{c}}{\~{a}}o de Aperfei{\c{c}}oamento de Pessoal de
N{\'{\i}}vel Superior (CAPES)  are gratefully acknowledged.

\end{document}